\documentstyle[prd,aps,preprint,eqsecnum,epsfig]{revtex}

\newcommand{\be}{\begin{equation}}
\newcommand{\ee}{\end{equation}}
\newcommand{\bea}{\begin{eqnarray}}
\newcommand{\eea}{\end{eqnarray}}

\newcommand{\bm}{\boldmath}
\newcommand{\fr}[2]{\frac{{\displaystyle #1}}{{\displaystyle #2}}}

\newcommand{\la}{\langle}
\newcommand{\ra}{\rangle}
\newcommand{\fn}[1]{\footnote{{\normalsize #1}}}

\newcommand{\epe}{\mbox{$e^+e^-$ }}
\newcommand{\ggam}{\mbox{$\gamma\gamma$ }}
\newcommand{\egam}{\mbox{$e\gamma$ }}

\def\lsi{\raise0.3ex\hbox{$<$\kern-0.75em\raise-1.1ex\hbox{$\sim$}}}
\def\gsi{\raise0.3ex\hbox{$>$\kern-0.75em\raise-1.1ex\hbox{$\sim$}}}
\newcommand{\lsim}{\mathop{\lsi}}
\newcommand{\gsim}{\mathop{\gsi}}

\newenvironment{Itemize}{\begin{list}{$\bullet$}%
{\setlength{\topsep}{0.2mm}\setlength{\partopsep}{0.2mm}%
\setlength{\itemsep}{0.2mm}\setlength{\parsep}{0.2mm}}}%
{\end{list}}
\newcounter{enumct}
\newenvironment{Enumerate}{\begin{list}{\arabic{enumct}.}%
{\usecounter{enumct}\setlength{\topsep}{0.2mm}%
\setlength{\partopsep}{0.2mm}\setlength{\itemsep}{0.2mm}%
\setlength{\parsep}{0.2mm}}}{\end{list}}

\title {The visible effect of a very heavy magnetic monopole at
colliders}

\author{I. F. Ginzburg\thanks{E-mail: ginzburg@math.nsc.ru}}
\address{Institute of Mathematics, Novosibirsk, 630090, Russia}

\author{A. Schiller\thanks{E-mail: schiller@tph204.physik.uni-leipzig.de}}
\address{Institut f\"ur Theoretische Physik and NTZ,
         Universit\"at Leipzig, D-04109 Leipzig, Germany}

\date{10 March 1999}

\tighten

\begin{document}
\draft
\preprint{UL--NTZ 09/99}

\maketitle

\begin{abstract}
If a heavy Dirac monopole exists, the light--to--light scattering below
the monopole production threshold is enhanced due to strong coupling
of monopoles to photons.  At the next Linear Collider with electron
beam energy 250 GeV this photon pair production could be observable
at monopole masses less than 2.5--6.4 TeV in the $e^+e^-$ mode or
3.7--10 TeV in the $\gamma\gamma$ mode, depending on the monopole
spin. At the upgraded Tevatron such an effect is expected to be
visible at monopole masses below 1--2.5 TeV.  The strong dependence
on the initial photon polarizations allows to find the monopole spin
in experiments at $e^+e^-$ and $\gamma\gamma$ colliders. We consider
the $Z\gamma$ production and the $3\gamma$ production at $e^+e^-$ and
$pp$ or $p\bar{p}$ colliders via the same monopole loop. The
possibility to discover these processes is significantly lower than
that of the $\gamma\gamma$ case.
\end{abstract}


\narrowtext

\section{Introduction}

The magnetic charge (monopole) was introduced by Dirac \cite{1} to
restore the symmetry among electricity and magnetism in particle
theory. Dyons --- particles with both magnetic and electric charge
--- were considered later by Schwinger \cite{2} (see also
\cite{dyo}). All attempts to discover the monopole gave
negative results, and there is no place for such a particle in modern
theories of our world. This is the reason, why the members of the
physics community do not believe in the existence of this particle.
However, there are no precise reasons against its existence.
Moreover, the idea of the monopole is very attractive: if it exists
somewhere, the mysterious quantization of the electric charge of
particles can be explained.

We discuss here only the {\bf point--like } monopole, assuming that it
was not observed yet only due to its very high mass. (This particle
differs strongly from nonlocal monopoles discussed in the context of
gauge theories after Polyakov and 't Hooft papers \cite{4}.)

Our basic idea is simple: {\em The existence of monopoles provides
for a $\gamma\gamma$ elastic scattering at large angles (``via
monopole loop''), which is sufficiently strong below the monopole
production threshold. This effect is observable at colliders with
energies smaller than the monopole mass. It can be used as method to
find a lower bound for the monopole mass from data.} The idea was
proposed for the first time in 1982 in Ref.~\cite{GPanf} where first
calculations for \epe and \ggam colliders were presented. The same
effect for hadron collider was considered recently \cite{Sch}. The
calculations of the last paper were the base to estimate a lower
limit for the monopole mass from D0 data \cite{land}. A resembling
idea for the process $e^+e^-\to Z\to 3\gamma$ has been used in
Refs.~\cite{Ruj,BBCL} and tested at LEP \cite{Acc}.  A similar
monopole loop effect should influence the photon and $Z$ boson
polarization operators. The last effect allows to obtain limitations
for the monopole mass from the anomalous magnetic momentum of the
muon \cite{muon} ($>120$ GeV) and data at the $Z$ peak \cite{Ruj}
($>1$ TeV).

In this paper we present more detailed calculations for the
production of large angle high energy photons at colliders signaling
the presence of virtual monopole loops and consider some new features
of this observable effect. We compare the discovery potential at
different future colliders and for several processes with monopole
loops. We show how to define the spin of the monopole via well
observed polarization effects at \epe or \ggam colliders. The
$3\gamma\,$ production at \epe and $p\bar{p}$ ($pp$) colliders via a
monopole loop is found to be less suitable for observing the effect.

Besides, we consider the similar $Z\gamma$ production.

Throughout the paper we denote the monopole mass by $M$ and its
spin by $J$ (we consider the opportunities $J=0$, $J=1/2$, or
$J=1$).

\section{Effective dynamics for heavy monopoles interacting with photons 
and $Z$ bosons much below the production threshold}
\label{secbasic}

\subsection{Basic assumptions}

A theory with two point--like charges, electric and magnetic,
cannot be standard QED. This theory needs to describe the
electromagnetic field with a vector potential having the Dirac
string \cite{1,2} or some its surrogate. To have unambiguous
results, the elementary electric and magnetic charges $e$ and
$g$ ought to be quantified so that
\be
g=\fr{2\pi n}{e}\quad n= \pm 1,\,\pm 2,\dots
\label{gquant}
\ee
with $\alpha \equiv e^2/(4\pi)=1/137 $ which leads to $\alpha_g\equiv
g^2/(4\pi) =n^2/(4\alpha)\approx 34n^2$.

The explicit form of such a theory is unknown (see Refs.
\cite{dyo,Ruj} and Appendix~\ref{appc}). Its construction needs
really new approaches. In particular, in the standard approach
this theory should even violate $U(1)$ local gauge invariance of
QED \cite{he}.

We consider below mainly the effective $4\gamma$ vertex, describing
in particular the light--to--light scattering.  It is assumed that the
theory of photon--monopole interactions restores the
electro--magnetic duality at least for an energy region below the
threshold of monopole production (the region of our interest)
\be
\omega\ll M\,.
\label{omegaregion}
\ee
Here $\omega$ is the characteristic photon energy in the discussed
process. The main features of a typical matrix element with external
photons in this region are:
\begin{Itemize}
\item gauge invariance provides a factor $\omega$ for each photon
leg;
\item to make this factor dimensionless it should be written   as
$\omega/M$;
\item a factor $g$ belongs to each vertex.
\end {Itemize}
Therefore, the amplitude with $n$ external photons
is at least proportional to $\left(g\omega/M \right)^n$. (In fact,
for
$n>4$ photons  an additional factor
$(\omega/M)^{n-4}$ arises --- cf. Ref.~\cite{HE}.)

We assume that at low enough photon energies the photon--monopole
interaction can be described effectively as a QED--like theory which
is valid in tree and 1--loop approximation with the expansion
parameter
\be
 g_{\text{eff}}=c(J)\fr{g}{4\pi}\fr{\omega}{M}\equiv
\fr{c(J)}{4\sqrt{\pi\alpha}}\fr{\omega n}{M}<1\,.
\label{region}
\ee
Here $c(J)$ is some numerical factor of order unity depending on the
monopole spin. Such a theory is gauge invariant and has the correct
low energy behavior. Besides, at $e/(4\pi) \ll g_{\text{eff}} < 1$
effects of ordinary particle production can be neglected.

We expect that with increasing $\omega$ (when our interaction becomes
formally strong) the real interaction strength of such a theory
becomes saturated. So, multiloop effects of QED, including essential
contributions from regions $|q_i^2|\gsim M^2$, with $q_i$ being
Euclidean loop momenta, are smoothed.  (From experience at strong
coupling is is known that cross sections are usually less than those
of low order perturbation theory.)

The validity range of the used approach is given by the following
limitations:
\begin{Enumerate}
\item multiloop radiative corrections should be small;
\item production cross sections for larger numbers of
photons should be small;
\item the unitarity for the elastic \ggam scattering should not be
violated.
\end{Enumerate}
Let us discuss the status of these limitations in more detail.

1) In standard QED the integration over loop momentum $q$ 
is convergent for the \ggam $\to$ \ggam process
due to gauge invariance.
Therefore, for loop integrals we can perform a Wick rotation into
the Euclidean region. In this case the integration region is limited
by virtualities $|q^2|\lsim\omega^2$, where the effective expansion
parameter is small, and our QED--like approach is valid. If some
convergent QFT--like monopole dynamics exists, these arguments show
that radiative corrections should be $\sim g_{\text{eff}}^2$ and
multiloop effects are negligible.

At a first glance, the experience with radiative corrections in QED
contradicts the above statement. The 2--loop calculations in standard
QED \cite{Ritus} give corrections $\sim \alpha/\pi$ which would
transform to meaningless large values by simply replacing $\alpha \to
\alpha_g$. However, these corrections have no relation to our
problem. Indeed, these $\omega$--independent corrections in QED arise
from integration regions with virtualities $|q_i^2|\sim m_e^2$ (i.e.
for our problem the inequality (\ref{omegaregion}) with $m_e\to M$ is
strongly violated).  Here the monopole dynamics should be quite
different from QED.  As it was noted above, this interaction is
expected to be more smooth than that of QED.

2) Based on the general Heisenberg--Euler effective Lagrangian for
heavy spin 1/2 and spin 0 particles in the loop \cite{HE,spin0}, it is
easy to see that the production of additional photons is suppressed
by a factor $(\omega/M)g_{\text{eff}}$ per photon.  This factor is
very small within the region (\ref{region}) (see e.g. \cite{Ruj}).

3) To respect the unitarity limit $\sigma_{\text{uni}}(\omega) \sim
4 \pi (2 J+1)/ \omega^2$ (the result for scalar particles is used,
for simplicity), the cross section of the considered \ggam $\to$
\ggam process has to be less than the sum of $S$ and $D$ partial
waves, describing the process at $\omega\ll M$ 
\be
\sigma_{\gamma\gamma\to\gamma\gamma}(\omega)<\fr{24 \pi}{\omega^2} \,.
\ee
In other words, our theory can be valid only at small enough energy
\be
\omega<\omega_m   \quad
\left[\sigma_{\gamma\gamma\to\gamma\gamma}(\omega_m)=24\pi/\omega^2_m\right]
\,.
\label{region1}
\ee
Using the cross section (\ref{sigtot1}) given below at
$\omega=\omega_m$, we obtain at $g_{\text{eff}}\approx1$ estimates
for $c(J)$.  Therefore inequalities (\ref{region}) and
(\ref{region1}) are satisfied simultaneously if $c(J=1)\sim 2.41$,
$c(J=1/2)\sim 1.33$, $c(J=0)\sim 0.94$.  These estimates well agree
with the experience in usual theories where we have for the effective
coupling constant $c(J=1)\approx 4$ (QCD), $c(J=1/2)\approx 2$ (QED),
$c(J=0)\approx 1$.

We conclude that {\em using one--loop results of QED without
radiative corrections within the region (\ref{region1}) is a
self--consistent approach, which has chances to be the correct
description of nature.}

\subsection{The 4--photon effective Lagrangian}

\begin{figure}[!htb]
     \begin{center}
      \leavevmode
      \epsfig{file=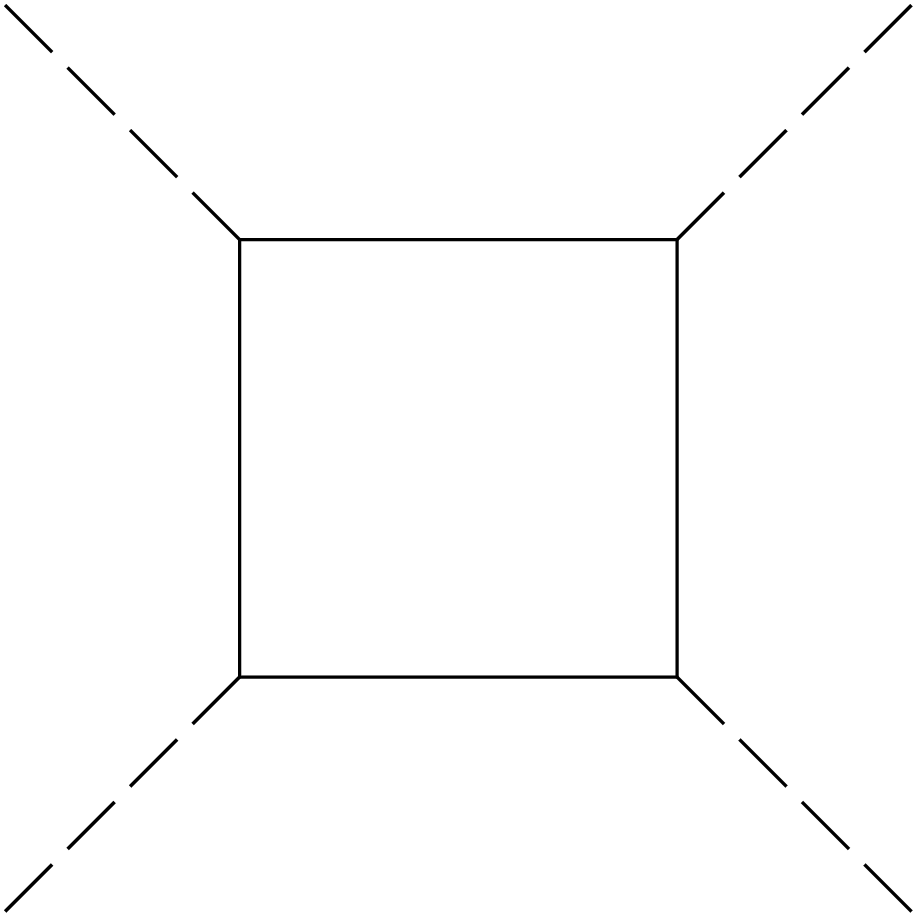,height=49mm,angle=270}
      \vspace{3mm}
      \caption{$\ggam \to \ggam$ via monopole loop}
      \label{fig:1}
    \end{center}
\end{figure}

Taking into account the above stated arguments we describe the
photon--monopole interaction by a Heisenberg--Euler effective
Lagrangian with coefficients given by standard QED.  For the
4--photon interaction via a monopole loop (Fig.~\ref{fig:1})
we have
\be
{\cal L}_{\text{eff}} = -\fr{F_{\mu \nu}F^{\mu \nu}}{4}
+ {\cal L}_{4\gamma}+\dots
\label{lagr}
\ee
with
\widetext
\be
{\cal L}_{4\gamma}=
 \fr{1}{36} \left(\fr{g}{\sqrt{4\pi} M}\right)^4
\left[\fr{\beta_++\beta_-}{2}
\left(F^{\mu\nu}F_{\mu\nu}\right)^2
+\fr{\beta_+-\beta_-}{2}\left(F^{\mu\nu}\tilde{F}_{\mu\nu}\right)^2\right]\,.
\label{lagr4g}
\ee
\narrowtext
Here $F_{\mu\nu}$ is the electromagnetic field strength tensor and
$\tilde{F}^{\mu \nu}= \epsilon^{\mu \nu \alpha\beta} F_{\alpha
\beta}/2$. (We assume also an additional gauge--fixing term to invert
the photon propagator.) The constants $\beta_\pm$ depend on the
monopole spin. The numerical coefficient is introduced to simplify
the numbers $\beta_\pm$. We use also the combinations
\be
P=\beta_+^2+2 \beta_-^2\,,\quad B=\fr{\beta_+^2-2 \beta_-^2}
{\beta_+^2+2\beta_-^2}\,.
\label{PB}
\ee
The numbers for these coefficients are collected in Table~\ref{t:1}.

The effective Lagrangian (\ref{lagr}) is valid to describe processes
with virtual photons having virtualities $|q^2|\lsim\omega^2$. The
relevant scale for the virtuality dependence of the \ggam $\to$ \ggam
cross section (possible form factor) is given by the unique inner
parameter of the monopole loop --- the monopole mass $M$.  This
dependence appears as ratios $|q^2|/M^2$ which are very small in the
region (\ref{region1}).  Therefore, the light--to--light cross section
can be used on-shell.

\subsection{Monopoles interacting with the $Z$ boson. Effective
Lagrangian}

With a heavy monopole the standard $SU(2)\times U(1)$ theory has to
be considered as low energy limit of some unknown theory. New
hypotheses are necessary to describe the interaction of monopoles
with $Z$ and $W$ bosons. We consider the simplest one
\cite{Ruj,BBCL}. For heavy monopoles ($M\gg M_Z$) it is naturally to
assume that the monopole interacts with the fundamental field of the
$SU(2)\times U(1)$ theory before symmetry breaking, namely, the
isoscalar field $B$.  Since the physical photon and the $Z$ boson are
obtained by rotation in the ($B,\,W^0$) plane, the monopole
interactions with $Z$ and $\gamma$ are assumed to be identical except
of an additional factor $\tan\theta_W$ in the monopole coupling to
$Z$.  In this approach the additional term for the $Z3\gamma$
coupling in the effective Lagrangian has the same form as in
Eq.~(\ref{lagr4g}) with the natural change of one field strength
$F_{\mu\nu}\to Z_{\mu\nu}$:
\widetext
\be
{\cal L}_{Z3\gamma} =
\fr{1}{9} \tan\theta_W  \left( \fr{g}{ \sqrt{4 \pi} M} \right)^4
\left[\fr{\beta_++\beta_-}{2}
\left(F^{\mu\nu}F_{\mu\nu}\right)\left(Z^{\mu\nu}F_{\mu\nu}\right)
+\fr{\beta_+-\beta_-}{2}\left(F^{\mu\nu}\tilde{F}_{\mu\nu}\right)
\left(Z^{\mu\nu}\tilde{F}_{\mu\nu}\right)\right].
\label{Zlagr}
\ee
\narrowtext
The factor of 4 in the definition of the $Z3\gamma$ coupling as
compared with Eq.~(\ref{lagr4g}) is chosen in such a way that the two
Lagrangians lead to the same strength, and the expression in momentum
space, for the $Z3\gamma$ and $4\gamma$ couplings modulo a factor
$\tan\theta_W$. For scalar or spinor monopoles the coefficients
$\beta_\pm$ in ${\cal L}_{Z3\gamma}$ are the same as those in ${\cal
L}_{4\gamma}$ (\ref{lagr4g}), a possible difference in $\beta_{\pm}$
for vector monopoles \cite{BBCL} is neglected in the following.

\section{Helicity amplitudes and cross sections}

\subsection{ Light to light scattering}

The helicity amplitudes corresponding to the effective Lagrangian
(\ref{lagr4g}) can be written in compact form ($\lambda_{1,2}$
--- helicities of initial photons, $\lambda_{3,4}$ --- helicities of
final photons)
\be
{{\cal M }}(\lambda_1, \lambda_2 ; \lambda_3 ,\lambda_4)=
\fr{8}{9}\left(\fr{g\omega}{\sqrt{4\pi}M}\right)^4 \;
{{\cal N }}(\lambda_1, \lambda_2 ; \lambda_3 ,\lambda_4)
\ee
with the only nonvanishing amplitudes
\[
{\cal N}(\lambda,\lambda; \lambda, \lambda)= 4 \beta_+ \,, 
\]
\be
{\cal N}(\lambda,\lambda; -\lambda, -\lambda)=2 \beta_- (3+\cos^2\theta)\,,     
\label{gamampl} 
\ee
\[
{\cal N}(\lambda,-\lambda; \pm \lambda, \mp \lambda)=
\beta_+ (1 \pm \cos \theta)^2 \,.
\] 
The corresponding total cross section of the $\gamma\gamma$ elastic
scattering is given by
\be 
\sigma(\gamma_{\lambda_1}\gamma_{\lambda_2}\to\gamma\gamma)
=\sigma_{\text{unp}}^{\text{tot}}
\left[1+\fr{5-2B}{7}\lambda_1\lambda_2\right]\,,
\label{sigtot}
\ee
\be
\sigma_{\text{unp}}^{\text{tot}}=R\omega^6 \,,  
\label{sigtot1}
\ee
\be
R = \fr{28P}{405\pi} \left(\fr{g}{\sqrt{4\pi} M}\right)^8\equiv
\fr{28P}{405\pi}\left(\fr{n}{2\sqrt{\alpha}M}\right)^8\,.
\label{sigtot2}
\ee

Let $s$, $t$ and $u$ be the standard Mandelstam variables
($s+t+u=0$), $\theta$ is the scattering angle in the c.m.s.  (with
$0<\theta<\pi/2$). In our case $s=4\omega^2$,\ $t =
-2\omega^2(1-\cos\theta)$.  The angular distribution is roughly
isotropic, but the details of the distribution are nontrivial:
\widetext
\bea
d\sigma(\gamma_{\lambda_1}\gamma_{\lambda_2}\to\gamma\gamma) &=&
\fr{5}{56}\sigma^{\text{tot}}_{\text{unp}}\left[
\left(3+\cos^2\theta\right)^2(1-B \lambda_1\lambda_2)
+8(1+B)\lambda_1\lambda_2\right]d\cos\theta  
\nonumber\\
&=&
\fr{5}{7}\sigma^{\text{tot}}_{\text{unp}}
\left[\left(\fr{s^2+t^2+u^2}{s^2}\right)^2
(1-B \lambda_1\lambda_2)+2(1+B)\lambda_1\lambda_2\right]\fr{dt}{s}
\,.
\label{sigdif}
\eea
\narrowtext 

\subsection{$Z \gamma$ production}

The helicity amplitudes for the process $\gamma\gamma\to Z \gamma$ are
obtained from the effective Lagrangian (\ref{Zlagr}):
\bea
{\cal M}(\lambda_1, \lambda_2 ; \lambda_3 ,\lambda_Z)&=&
 \fr{8}{9} \tan\theta_W \left(\fr{g\omega}{\sqrt{4\pi}M}\right)^4 \sqrt{\rho} 
 \nonumber \\
  &&\times {\cal N}(\lambda_1, \lambda_2 ; \lambda_3 ,\lambda_Z)
\eea
with $\rho =1-M_Z^2/s$, $\lambda_Z$ is the helicity of the $Z$. 
In addition to the helicity amplitudes (\ref{gamampl}), 
the further nonvanishing amplitudes are 
\[
{\cal N}(\lambda,\lambda; -\lambda ,\lambda)= 
2 \rho \; \beta_- \sin^2\theta \,, 
\]
\be
{\cal N}(\lambda, \lambda; -\lambda, 0)=
2 \sqrt{2 \rho} \; \beta_-  \lambda \sin\theta \cos\theta \,,
\label{Zampl}
\ee
\[
{\cal N}(\lambda,-\lambda;\pm \lambda,\pm\lambda)=
\rho \; \beta_+   \sin^2\theta \,,
\]
\[
{\cal N}(\lambda, -\lambda; \pm \lambda, 0)=
\sqrt{2 \rho} \; \beta_+   \lambda (1\pm \cos\theta) \sin \theta\,.
\]

At large energies ($s\gg M_Z^2$) we obtain the angular distribution
including the polarizations of initial photons:
\widetext
\bea
d\sigma(\gamma_{\lambda_1}\gamma_{\lambda_2}\to Z\gamma)&=&
\fr{2}{81 \pi} \tan^2\theta_W \fr{1}{\omega^2}
\left(\fr{g\omega}{\sqrt{4\pi} M}\right)^8
\left\{\beta_+^2(3+\cos^2\theta)+\beta_-^2(5+3\cos^2\theta)\right.
\nonumber\\
&&\left.+\lambda_1\lambda_2\left[\beta_+^2(1-\cos^2\theta)+\beta_-^2
(5+3\cos^2\theta)\right]\right\} d \cos \theta \,. 
\label{zgamdifsec}
\eea
Using the notations of Eqs.~(\ref{PB}),(\ref{sigtot1}) the
corresponding total cross section can be written as
\be
\sigma(\gamma_{\lambda_1}\gamma_{\lambda_2} \to Z\gamma)=
\fr{5}{42}\tan^2\theta_W[(19 + B)+\lambda_1\lambda_2
(11- 7 B)]\sigma^{\text{tot}}_{\text{unp}}(\gamma\gamma\to
\gamma\gamma)\,.
\label{zgamcrsec}
\ee
\narrowtext
This cross section is less than that of the \ggam $\to$ \ggam process
(\ref{sigtot1}). With the lower detection efficiency for the $Z$ boson
compared with the high energy photon, the process $\gamma\gamma\to
Z\gamma$ is less suitable than the light--to--light scattering to
discover the monopole loop effect.
 
\subsection{$3\gamma$ production. Total cross section}

\begin{figure}[!htb]
     \begin{center}
      \leavevmode
      \epsfig{file=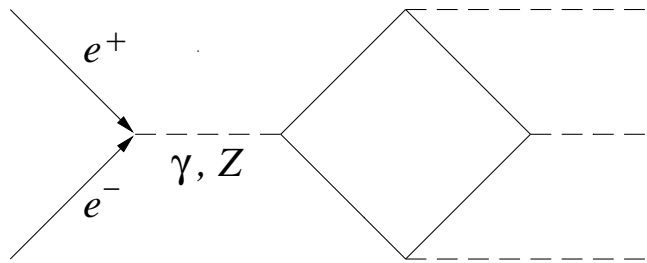,height=85mm,angle=270}
      \vspace{3mm}
      \caption{$e^+e^- \to 3 \gamma$ via monopole loop}
      \label{fig:2}
    \end{center}
\end{figure}

We consider the $3\gamma$ production in \epe or $q\bar{q}$
collisions.  Let us denote the total cross section of the process
\epe $\to 3\gamma$ via a monopole loop (Fig.~\ref{fig:2}) with a
$s$--channel photon by $\sigma_\gamma(3\gamma)$.  In the total cross
section both $Z$ and $\gamma$ exchange have to be taken into account.
The monopole loop factors are identical for both exchanges up to
$\tan\theta_W$ in the amplitude (coefficents $\beta_\pm$ for $J=1$
monopoles in $4\gamma$ and $Z3\gamma$ couplings are set to be equal).
With these changes we have at $s\gg M_Z^2$ for \epe or $q\bar{q}$
collisions
\be
\sigma(f^i\bar{f}^i\to 3\gamma) =K_i \sigma_\gamma(3\gamma)\,, 
\label{11}
\ee
\[
K_i=
\left(q_i+\fr{g_V^i}{2\cos^2\theta_W}\right)^2 +
\left(\fr{g_A^i}{2\cos^2\theta_W}\right)^2\,.
\]
Here $g_V^i$ and $g_A^i$ are the vector and axial coupling of fermion
$f^i$, $q_i$ is the charge of $f_i$ in units of $e$.  For electrons
we obtain $K_e=1.174$, for $u$--quarks and $d$--quarks $K_u=0.739$
and $K_d=0.416$, respectively.  Using the results of Ref.~\cite{BBCL}
for the process \epe $\to Z \to 3\gamma$ via a monopole loop, we
write the cross section $\sigma_\gamma(3\gamma)$ (for the $e^+$ and
$e^-$ energy $E$) in the notations for the
\ggam $\to$ \ggam process given in Eq.~(\ref{sigtot2}) as
\be
\sigma_\gamma(3\gamma) = \fr{\alpha}{252\pi} RE^6
 \fr{3\beta_+^2 +5\beta_-^2}{3\beta_+^2
+6\beta_-^2} \,.
\label{sig3gam}
\ee
Differential distributions for the $3\gamma$ production in \epe
collisions via $Z$ exchange were analyzed in Ref.~\cite{BBCL}. The
result at the $Z$ peak was given in Ref.~\cite{Ruj}.

\section{Discovering the monopole effect at colliders}

Let us discuss the opportunity to see the effect described by the
effective Lagrangian (\ref{lagr4g}) at \ggam, \epe and proton
colliders.  We focus our interest to a first observation where the
cross section is not very large and the collider energy is $E\ll M$.
In this region the cross section of the process raises rapidly with
the energy of the initial (real or virtual) photons.  So, the effect
will be seen as the production of high energy photons at large
angles.

\subsection{Photon Colliders}

The constructed linear colliders will be used both in \epe mode
and in \ggam or \egam modes  ({\bf Photon Colliders}) with
the following typical parameters ({\em obtained without a special
optimization for the photon mode}) \cite{GKST,SLACDESY}.

\begin{Itemize}
\item {\em Characteristic photon energy $\omega\approx 0.8E$.}
($E=0.25\div 1$ TeV is the electron energy in the basic $e^+e^-$
collider).
{\em
\item Annual luminosity ${\cal L}_{b}\approx 100$
{\rm fb}$^{-1}$ (or larger).
\item Mean energy spread $\langle\Delta\omega\rangle  \approx 0.07\omega$.
 \item Mean photon helicity $\langle\lambda_\gamma\rangle
\approx 0.95$, the sign of which can be changed easily
{\rm \cite{GKST}}.
}
\end{Itemize}

The considered process will be seen as elastic large angle light--to--light 
scattering. Since the cross section raises quickly with energy,
our effect will be most easily observed for photons with highest
energy.  The scattering at $\theta=\pi/2$ is almost twice as large as
the forward scattering (see Eq.~\ref{sigdif}).  
Both total and differential cross sections depend
strongly on the sign of the initial photon helicities.
This dependence is very
different for different values of the monopole spin.

{\bf The background} is given by the light--to--light scattering via a
W--boson loop with a cross section of about 20 fb at the discussed
energies\cite{Jik}.\fn{The numerical results of that paper should be
reduced by a factor $0.76=(128/137)^4$ since the photons here are
real, and the fine structure constant is $\alpha=1/137$, but not
$1/128$.} In contrast to the effect of interest, the background
angular distribution of the produced photons is shifted to the beam
collision axis, with transverse momenta $\lsim M_W$. Besides, this
background varies slowly with energy.

Neglecting the difference in the angular distributions for signal and
background at 0.25--1 TeV, we conclude: To see our effect
unambiguously with a good Signal/Background ratio, a production cross
section above 1 fb is necessary.

\subsection{\bm\epe Colliders}

To search for the monopole effect in \epe collisions, the emission of
high energy photons with large transverse momenta via photon fusion
should be studied (\epe $\to$ \epe \ggam). To describe this process
we use the leading log equivalent--photon approximation (cf. e.g.
\cite{9}) having an accuracy $\sim 1/\ln(E^2/m_e^2)\approx 0.05$.

Let us denote  the energies of the separate virtual photons by 
$\omega_i=x_iE$. Then the effective mass of produced \ggam system is
$\hat{s}=4\omega_1\omega_2\equiv 4x_1x_2E^2$. The result for the
collision of unpolarized electrons is written with high accuracy via
the spectra of the equivalent photons $dn_i$ ($Q^2=-q^2>0$):
\be
\sigma(ee\to ee\gamma\gamma)=RE^6
\int dn(x_1)dn(x_2)(x_1x_2)^3\,, 
\label{6}
\ee
\be
dn(x)=\fr{\alpha}{ \pi} \fr{dx}{x}
\int\limits_{Q^2_{\text{min}}}^{Q^2_{\text{max}}}
\fr{dQ^2}{ Q^2}\left[1-x +\fr{1}{2} x^2
-(1-x)\fr{Q^2_{\text{min}}}{Q^2}
\right] \,,   
\label{6a}
\ee
\[
Q^2_{\text{min}}= \fr{m_e^2x^2}{1-x}\,.
\]
Since the virtuality dependence for the subprocess cross section is
negligible, the integration over the virtuality is spread over the
whole kinematical region, till $Q^2_{\text{max}}\approx \hat{s}$,
where the product of the photon fluxes in Eq.~(\ref{6}) become much
higher than the precise QED expression.  Within the
considered accuracy we obtain from Eq.~(\ref{6a}) the
photon spectra
\be
dn(x_i)=\fr{\alpha}{\pi} L \;f(x_i)\,, 
\label{spectr} 
\ee
\[
f(x_i)=1-x_i+\fr{1}{2} x_i^2  \,, \quad 
L=\ln\fr{4E^2}{m_e^2} \,.
\]

To take into account polarization effects of initial leptons in
Eq.~(\ref{6}) we note that the doubled longitudinal electron and
positron helicities $\xi_i$ are transfered to the mean helicities of
their equivalent (virtual) photons $\lambda_i^v$ via \cite{GinSer}
\be
\lambda^v =A(x)\, \xi \,,\quad A(x)=\fr{x-x^2/2}{1-x+x^2/2}\,.
\ee
Therefore, the integrand in the total cross section (\ref{6})
acquires the additional factor
\be
\left[1+\fr{5-2B}{7}A(x_1)A(x_2)\,\xi_1\xi_2\right].
\label{epepol}
\ee

Combining Eqs.~(\ref{6})-(\ref{epepol}), the energy spectrum for the
initial equivalent photons takes the form
\widetext
\be 
d\sigma(e_{\xi_1}e_{\xi_2}\to ee\gamma\gamma)=   
\fr{\alpha^2}{\pi^2} RE^6 L^2
\left[ 1+\fr{5-2 B}{7} A(x_1) A(x_2)\, \xi_1\xi_2\right]
 f(x_1) x_1^2 dx_1 f(x_2) x_2^2 dx_2 \,,
\label{LLA}
\ee
\narrowtext
and we calculate the total cross section, the characteristic photon
energy $\la\omega\ra$ and the mean spread of the virtual
photon energy $\la\Delta\omega\ra$\footnote{Eqs.~(\ref{8a}), (\ref{8b}) were obtained in
Ref.~\cite{GPanf}.}:
\be 
\sigma(e_{\xi_1}e_{\xi_2}\to ee\gamma\gamma)=R\alpha^2E^6D^2
\left[1 + \fr{81}{847} \left(5-2B\right) \xi_1 \xi_2\right]\,, 
\label{sig2g} 
\ee
\be
D=\fr{1}{ \alpha E^3}\int\limits^E_0  \omega^3dn \approx
\fr{11}{60\pi}L\,,
\label{7} 
\ee
\be
\langle\omega\rangle=\fr{\int\omega\times\omega^3 dn}
{\int \omega^3dn}=
\fr{8}{ 11}E\approx 0.73 E \,, 
\label{8a}  
\ee
\be
\la\Delta\omega\ra =\sqrt{\la \omega^2\ra -\la\omega\ra^2}=
\sqrt{\fr{36}{847}}E =0.283 \, \la\omega\ra \,.
\label{8b}
\ee
So, the effect looks like the production of a resonance with
$M\approx 1.4E$ and $\Gamma/M\approx 0.283$.  The ``resonance'' will
be produced almost at rest. For the considered energies $E=100-1000$
GeV we have $D=1.50-1.77$.  Differential cross sections for the
produced photons are given in Appendix~\ref{appa}.
 
A different possibility for observing the influence of the monopole
loop effect seems to be the production, through annihilation, of
three photons (see Fig.~\ref{fig:2}). In the discussed energy range
the ratio of total cross sections for three to two photon production
(Eqs.~(\ref{11}) and (\ref{sig2g})) is roughly $K_e/(252\pi\alpha
D^2)\approx 1/12$.  Therefore, the three photon production is
unsuitable for our goal.

{\bf Background}. A characteristic feature of the discussed process
is the radiation of two noncollinear high energy photons with
$\omega\sim E$. There are only two processes of standard QED with
radiation of two high energy photons having large $p_\perp$ and with
not too small cross sections.  The first of them is the process
\epe $\to$ \ggam with a cross section for large angle production
$\sim\alpha^2/E^2$. This process differs from that considered above
since here the photons are collinear and their energies $\omega=E$.
The second QED process is  \epe $\to 3\gamma$ with a very low cross
section for the radiation of large angle high energy photons ($\sim
\alpha^3/E^2$).

\subsection{Numerical estimates for \epe and \ggam colliders}
\label{secest}

The effects discussed have a very good signature and a small
background. Therefore, a relatively low cross section can be used to
demonstrate the observability of the effect.

Let us first summarize our estimates for the characteristic photon
energy $\langle\omega\rangle$ and the ratio of cross sections with
initial antiparallel and parallel helicities of photons or electrons
$\sigma_-/\sigma_+$ for the electron beam energy $E$ in
Table~\ref{t:2}. These ratios show the very high potential to
determine the monopole spin from measurements with polarized beams.

To estimate the possible monopole mass bound, we demand to have 10
events in the \epe mode of TESLA (at $E=250$ GeV and $E=1$ TeV with
luminosity integral 500 fb$^{-1}$) and a 1 fb cross section in the
\ggam mode of that collider (with 100 fb$^{-1}$ integrated
luminosity).  Table~\ref{t:3} shows the ultimate values of the
monopole mass (in TeV) depending on its spin at different initial
energies of the electron bunch $E$ (in TeV) and the corresponding
values of $\langle\omega\rangle/\omega_m$. For a possible experiment
at LEP2 we use the luminosity integral 0.5 fb$^{-1}$ and assume to
have 10 events. Besides, we include in Table~\ref{t:3} ``universal
estimates'' assuming that the process with a monopole loop is
observable for a cross section $\sigma\ge \sigma_b =0.1\alpha^2/(3E^2)$
for both \epe and \ggam collisions. This leads to $\sigma_b=0.7$ fb
at $E=1$ TeV and $\sigma_b=11$ fb at $E=0.25$ TeV.

Table~\ref{t:3} shows that we are far from the boundary
(\ref{region1}). Therefore, we expect no new phenomena with
Increasing energy except a fast rising cross section (for
example, the cross section becomes twice as large increasing the
collider energy by $12\%$).

\subsection{Proton colliders}

The proton colliders are those with highest beam energies.  The main
observable effect here should be $pp\to \ggam+\dots\,$ in the
collision of two virtual photons, created by the protons (or
antiprotons) in both elastic and inelastic processes ($p\to
p\gamma^*$, $p\to\gamma^*+...$) with the $\gamma^*\gamma^*\to \ggam$
subprocess.  These calculations were performed for the Tevatron and
the LHC in Ref.~\cite{Sch}, where total and differential cross
sections for the produced photons were obtained. The total cross
section (obtained numerically using known approximations for the
proton structure functions \cite{GRV} and form factors) can be
written in the form
\be
  \sigma_{pp \to \gamma\gamma X}
   = 108 \, P \, \left(\fr{n E}{M}\right)^8
   \left( \fr{N(E) }{N(1 \,{\text {TeV}}) } \right)^2 
   \left( \fr{ 1\,{\text {TeV}}}{  E } \right)^2
     {\text {fb}}\,.\label{ppcrs}
\ee
Here $N(E)$ is the normalized flux of the effective photons, it
varies within 10\% in the energy interval 1--7 TeV.
 
This relatively low (in comparison with the \epe case) cross
section is due to the soft structure of the proton with respect to
the photons. This fact results also in a relatively low fraction
of the total energy transfered to the photons.

With a good accuracy for the 1--7 TeV energy interval the average
energy of the colliding (virtual) photons and their energy spread are
\be
      \langle \omega \rangle =0.314 \, E\,,\quad
      \langle \Delta \omega \rangle =0.149 \, E \,.
\label{aomega}
\ee
The energies of produced photons are close to this value.

Luminosity integrals of 2 fb$^{-1}$ at $E=0.9$ TeV (upgraded
Tevatron) and 100 fb$^{-1}$ at $E=7$ TeV (LHC) have been considered.
Supposing 10 events to record the effect, the following ultimate
values for the monopole mass (in TeV) and corresponding values of the
ratio $\langle\omega\rangle/\omega_m$ were obtained \cite{Sch}, see
Table~\ref{t:4}.  The third line in this Table represents the results
of data processing at the Tevatron \cite{land} based on equations
from Ref.~\cite{Sch}. Here a much smaller luminosity integral 70
pb$^{-1}$ was used, and the limitations for the monopole mass are
much less restrictive.

To complete the study for proton colliders one has to estimate the
process $pp\to 3\gamma+...$ via a monopole loop arising from the $q
\bar q$ fusion into a single photon similar to Fig.~\ref{fig:2}.
We calculate the corresponding cross section, using
Eqs.~(\ref{11}),~(\ref{sig3gam}) and the GRV parametrizations for
the structure functions \cite{GRV}. It is useful to write the result
in the form
\bea
  \sigma_{pp \to \gamma\gamma\gamma X}
 &=& D(E)P \, \left(\fr{n E}{M}\right)^8
\left(\fr{3\beta_+^2+5\beta_-^2}{3\beta_+^2+6\beta_-^2}\right)
\nonumber \\
&&\times
\left( \fr{ 1\,{\text {TeV}}}{  E } \right)^2
     {\text {fb}}\,,
\label{pp3gamma}
\eea
\[
D({\text{Tevatron}}) = 4.2\,,\quad D({\text{LHC}})=0.17\,.
\]
The large difference in the cross sections for the Tevatron and the
LHC is due to the fact that the number of antiquarks in protons is
much lower than that in antiprotons at relatively large values of
the (anti)quark momentum fractions $x_i$.
 
These cross section are much smaller than those for the two--photon
case (\ref{ppcrs}). Therefore, this process is unsuitable to discover
the monopole effect.

\section{Final comments}

We have studied different processes suitable to discover the
point--like Dirac magnetic monopole in experiments at colliders
before its direct observation.  We have found that the study of the
two--photon final state produced by two initial photons either real
or virtual has the highest discovery potential. Among different
colliders the most promising ones are Next Linear Colliders in the
Photon Collider mode.  For both the Photon (\ggam) and \epe mode of
these colliders the ratio $\langle\omega\rangle/\omega_m\,$ is small
enough to neglect unwanted corrections in the theory.  With the lower
detection efficiency for $Z$ bosons compared with high energy
photon, studies of the $Z\gamma$ final state are less favorable.

For \epe and proton colliders the cross sections for the
$3\gamma$ final state via an intermediate photon or $Z$ state
(Fig.~\ref{fig:2}) is much smaller than that of the $\ggam$
production, they are unsuitable for a first observation of effects
caused by monopoles.

The same approach provides the opportunity to study processes of
higher order \ggam $\to 4\gamma,\dots$,  \epe $\to 5\gamma,\dots\,$  Their
cross sections are relatively small within the region of interest.
The increasing photon multiplicity due to these processes can be
observed only for an energy much larger than that necessary for a
first observation of the effect.

We would like to mention that also a monopolium bound state $R$ with
mass $M_R$ and spin $J_R$ can exist, as has been discussed, for
example, in Refs.~\cite{GPanf,Ruj}.  To make a rough estimate, we
assume $2 \omega \sim M_R$. If $M_R\gsim 8\pi M/(gc(J)) \approx 1.32M/(nc(J))$,
the monopolium resonance is invisible at the discussed energies (when
inequality (\ref{region1}) is valid).  However, there exists another
possibility: $M_R<8\pi M/(gc(J))\approx 1.32M/(nc(J))$ where the
effective coupling $g_{\text{eff}}$ is small near this resonance.
In this case the two--photon decay of $R$ is dominant, and (for a
C--even resonance with $J_R=0,2$) the cross section for the
transition \ggam $\to R\to$ \ggam with $2\omega\sim M_R$ is large.
Then at $2\omega>M_R$ a resonance peak could be expected in the
\ggam elastic scattering. The C--odd resonance with $J_R=1$ should be
observable as some new $Z$--like heavy vector boson decaying mainly
into three photons.

\acknowledgments

We thank G. B\'elanger and F. Boudjema for contributing to early stages
of this work.
We are very grateful to F.~Goldhaber, P.~Grannis and H.-J.~He for
useful discussions. 
{\em The work is supported by grants RFBR 99-02-17211
(I.F.G.) and Volkswagen Stiftung I\hspace{0.5mm}/72 302 (I.F.G. and
A.S.)}.

\appendix
\section{The energy and momentum distribution for the
produced photons at \epe collider}
\label{appa}

Since the produced \ggam system is almost at rest, we will have a
roughly isotropic angular distribution in \epe collisions.  More
detailed distributions can be obtained, substituting
Eq.~(\ref{sigdif}) into the integrand (\ref{6}) and performing a
subsequent boost and a space rotation. In a first approximation we
can neglect the transverse motion of the virtual photons (let us
remind that in the main part of the integration region we have
$Q_i^2\approx{\mathbf{q}}_{\perp i}^2/ (1-x_i)\ll \hat s$ with the
tranverse momenta of the virtual photons ${\mathbf{q}}_{\perp i}$).
Therefore, the transverse momenta of the produced photons are
balanced: $p_{\perp 3}\approx -p_{\perp 4}\equiv p_\perp$. Their
4--momenta in the c.m.s. of the protons are
\be
    p_{3,4}= p_\perp(\cosh\eta_{3,4},\pm 1,0,\sinh\eta_{3,4})
    \label{kinem}
\ee
with the rapidities of the produced photons $\eta_{3,4}$.
Using these notations we have $\sin\theta=p_\perp/(E\sqrt{x_1x_2})$
and $x_{1,2}= (p_\perp/2E) \left(\exp(\pm\eta_3)+\exp(\pm\eta_4)\right)$.
With the standard transformation
\[
\fr{\partial^3}{E^2\partial x_1\partial x_2
\partial \cos\theta}
  \equiv 2 \fr{\partial^3}{\partial p_\perp^2 \partial\eta_3 \partial\eta_4}
\]
the integrand of Eq.~(\ref{6}) with Eq.~(\ref{spectr}) represents the
transverse momentum--rapidity distribution of the produced photons.
The differential cross section of the $\ggam$ production (after
integrating over one transverse momentum and azimuthal angle) can be
written in the form
\be
\fr{d^3\sigma}{d\eta_3d\eta_4dp_\perp^2}= 
\fr{5 \alpha^2}{112 \pi^2} RE^4L^2 \Psi
f(x_1) x_1^2 f(x_2) x_2^2\,,
\ee
\[ 
\Psi=\Phi\left[1-BA(x_1)A(x_2)\xi_1\xi_2\right] 
+8(1+B)A(x_1)A(x_2)\xi_1\xi_2\,,
\]
\[
  \Phi =\left\{-\fr{p_\perp^2}{E^2x_1x_2}\right\}^2\equiv
  \left\{4- \fr{1}
  {\cosh^2\left[\left(\eta_3-\eta_4\right)/2\right]}\right\}^2 \,.
\]  
 
Thus the measurement of this differential cross section allows to
determine the monopole spin via the polarization dependence of the
distribution.

{\bf $\gamma\gamma$ total transverse momentum distribution.} 
The total transverse momentum of the produced photon pair ${\mathbf{k}}_\perp
\equiv {\mathbf{p}}_{\perp 3}+{\mathbf{p}}_{\perp 4}$ is equal to the
sum of transverse momenta of the virtual photons,
${\mathbf{k}}_\perp={\mathbf{q}}_ {\perp 1}+ {\mathbf{q}}_{\perp 2}$.
This momentum  distribution (written with logarithmic
accuracy) can be found in Ref.~\cite{9} (we use Eq.~(5.31) from that
paper restoring omitted terms $\sim x_i^2$ in the photon spectra). In our
case the mean value of $x_i$ is about 0.7.  Therefore, with the used
accuracy we also neglect these factors in the argument of the logarithm.
Taking into account the electron polarization
we have
\widetext
\be 
d\sigma(e_{\xi_1}e_{\xi_2} \to ee \gamma\gamma) =\fr{2\alpha^2}{\pi^2}RE^6
\left[1+\fr{5-2B}{7}A(x_1)A(x_2)\xi_1\xi_2\right]
\fr{d{\mathbf k}_\perp^2}{{\mathbf k}_\perp^2}
\ln \fr{{\mathbf k}_\perp^2}{m_e^2}  f(x_1) x_1^2 dx_1 f(x_2) x_2^2 dx_2
\label{ktx1x2}
\ee
and
\be
d\sigma(e_{\xi_1}e_{\xi_2} \to ee \gamma\gamma) =
\fr{121\alpha^2}{1800\pi^2}RE^6
\left[1+\fr{81}{847} \left(5-2B\right) \xi_1\xi_2\right]
\fr{d{\mathbf k}_\perp^2}{{\mathbf k}_\perp^2}
\ln \fr{{\mathbf k}_\perp^2}{m_e^2}\,.
\label{ktdistr}
\ee
\narrowtext
After integrating Eq.~(\ref{ktx1x2})
over the total transverse momentum, the energy
spectrum coincides in leading logarithmic accuracy with
Eq.~(\ref{LLA}).

\section{Comments on possible calculations using string dependent
monopole--photon vertex}
\label{appc}

The string--dependent monopole--photon coupling vertex can be written
as (for spin 1/2 monopole)
\be
\Gamma_\mu(q)=ig\fr{\varepsilon_{\mu\nu\sigma\tau}n^\nu
q^\sigma\gamma^\tau}{nq-i\epsilon}\,.\label{string}
\ee
Here $q$ is the photon momentum and $n$ is a space--like unit vector
directed along the string (it is assumed to be a straight line for
simplicity). The choice of the string is arbitrary: a reorientation
of the string is a kind of gauge transformation.

Attempts to perform calculations using this vertex in the spirit of
standard QED gave no satisfactory results.  It seems that in the
``good'' theory some extra terms should be added to the interaction
(\ref{string}).  The known ``string independent'' result for dyon
scattering is obtained using the unjustified extra prescription to
consider only the most singular term $1/q^2$ in the amplitude at
$q^2\to 0$ (omitting the string piece in the result)
\cite{Schw2,Deans}.

The authors of Ref.~\cite{critiq} pretend for a solution of this
problem at least for  the light--to--light scattering.
Unfortunately, this proposal cannot be realized without strong
additional ideas which seem unknown till now.

Indeed, using this singular coupling the result depends on the
(nonphysical) direction of the string line $n$. Such a result seems
meaningless. In particular, the discussed cross section is infinite
for  photon momenta orthogonal to the string direction.

To avoid this difficulty one can try to average the reaction
amplitude over the directions of $n$ (using either different vectors
for different vertices or a common vector for all vertices). To avoid
technical difficulties, this averaging can be performed in a specific
frame where $n_0=0$ (e.g., the c.m.s.  of the light--to--light
scattering) \cite{recent}.

Unfortunately, this idea gives also no satisfactory results.  Indeed,
if a separate direction of $n$ for each vertex is considered, the
averaged vertex will be 0 (this can be concluded, e.g. from the
calculations of paper \cite{recent} by inserting their Eq. (14) into
Eq.~(\ref{string})).  Considering a common vector $n$ for all
vertices, it is easy to see that the cross section is divergent (due
to the singular nature of basic interaction (\ref{string})).
Therefore, both ``natural'' possibilities seem to be senseless.

There are no grounds for a difference between an electron and a
monopole in the validity range of the duality approach.
Introducing an additional factor $\omega/M$ for each photon leg as it
has been proposed in Ref.~\cite{critiq} results in a strong
difference between electricity and magnetism.  It breaks down the
basic idea of introducing a monopole. That is a new hypothesis which
is one among a large variety of unjustified assumptions.

\begin{table}[!htb]
\begin{center}
\begin{tabular}{ccccdd} 
$J$&Ref.&$\beta_+$&$\beta_-$&$P$&$B$\\ \hline
0 &\cite{spin0}&$1/5$&$3/20$&0.085&$-$0.059\\ \hline
1/2& \cite{HE}&$11/10$& $-3/10$&1.39&0.74\\ \hline
1 &\cite{BB,Jik}&$63/5$&$9/20$ &159.165&0.995\\  
\end{tabular}
\caption{Coefficients $\beta_\pm$, $P$ and B$$ for monopoles of
different spin}
\label{t:1}
\end{center}
\end{table}

\begin{table}[!htb]
\begin{center}
\begin{tabular}{cdddd} 
collider&$\langle\omega\rangle$&$J=0$&$J=1/2$&$J=1$\\\hline
\ggam &0.8$E$&0.155&0.33&0.40\\ \hline
\epe &0.73$E$&0.44&0.58&0.63\\  
\end{tabular}
\caption{Mean photon energy and ratio of cross sections with
initial antiparallel and parallel helicities of photons or electrons
$\sigma_-/\sigma_+$.}
\label{t:2}
\end{center}
\end{table}

\begin{table}
\begin{center}
\begin{tabular}{cccccd} 
$E$&collider&$J=0$&$J=1/2$&
$J=1$&${\langle\omega\rangle}/{\omega_m}$\\ \hline
0.25&\ggam &3.75$n$ &5.6$n$ & 10$n$& 0.078\\
\hline
0.25&\epe& 2.5$n$& 3.6$n$& 6.4$n$& 0.112\\ \hline
1.0&\ggam &11$n$ &16$n$ &28$n$& 0.110\\
\hline
1.0 &\epe& 7$n$&10$n$&18$n$& 0.154\\ \hline
0.1&LEP2&0.54$n$&0.76$n$&1.35$n$&0.21\\\hline
univ.&\ggam&11.8$nE$&16.75$nE$&30$nE$&0.105\\
univ.&\epe&4.6$nE$&6.5$nE$&11.7$nE$&0.25\\ 
\end{tabular}
\caption{Lower limits for monopole masses in TeV for \epe and
\ggam colliders depending on the collider energy in
TeV and the monopole spin and the ratio $\langle\omega\rangle/\omega_m$.}
\label{t:3}
\end{center}
\end{table}

\begin{table}[!htb]
\begin{center}
\begin{tabular}{cdddd} 
collider&$J=0$&$J=1/2$&
$J=1$&${\langle\omega\rangle}/{\omega_m}$\\ \hline
Tevatron(2 fb$^{-1})$&0.998$n$&1.42$n$&2.56$n$&0.44\\\hline
LHC&7.40$n$&10.5$n$&19.0$n$&0.45\\\hline
Tevatron\cite{land}&0.61$n$&0.87$n$&1.58$n$&0.63\\  
\end{tabular}
\caption{Same as in Table~\protect\ref{t:3} for proton colliders.}
\label{t:4}
\end{center}
\end{table}

\end{document}